\documentstyle[prl,aps]{revtex}

\font\Bbb =msbm10

\def\spec{R_{\alpha\beta}}
\def\id {{\hbox{\Bbb I}}}
\def\kvec{{\bbox{k}}}
\def\lvec{{\bbox{l}}}
\def\duzomniejsze{<\kern-.7mm<}
\def\duzowieksze{>\kern-.7mm>}
\def\intlarge{\mathop{\int}\limits}
\def\textbf#1{{\bf #1}}
\def\be{\begin{equation}}
\def\ee{\end{equation}}
\def\ben{\begin{eqnarray}}
\def\een{\end{eqnarray}}\def\be{\begin{equation}}
\def\ee{\end{equation}}

\def\eea{\end{array}}
\def\bea{\begin{array}}
\newcommand{\bei}{\begin{itemize}}
\newcommand{\eei}{\end{itemize}}
\def\ra{\rangle}
\def\la{\langle}

\def\lcal{{\cal L}}

\def\trace{\mbox{Tr}}
\def\dt#1{{{\kern -.0mm\rm d}}#1\,}
\def\ypodpis{\raise4mm\hbox{$\omega$}}

\begin{document}
\draft
\twocolumn

\title{Dynamical description of quantum computing: generic nonlocality
of quantum noise}

\author{Robert Alicki$^{1}$, Micha\l{} Horodecki$^{1}$,
Pawe\l{} Horodecki$^{2}$ and Ryszard Horodecki$^{1}$}

\address{$^1$ Institute of Theoretical Physics and Astrophysics,
University of Gda\'nsk, 80--952 Gda\'nsk, Poland,\\
$^2$Faculty of Applied Physics and Mathematics,
Technical University of Gda\'nsk, 80--952 Gda\'nsk, Poland
}

\maketitle

\begin{abstract}
We develop dynamical non-Markovian description of quantum computing in weak
coupling limit, in lowest order approximation. We show that the 
long-range memory 
of quantum reservoir (such as the $1/t^4$ one exhibited by electromagnetic vacuum)
produces strong interrelation between structure of noise
 and quantum algorithm, implying nonlocal attacks of noise.  This
shows that the implicit assumption of quantum error correction theory --
independence of noise and self-dynamics  --  fails in long time regimes.
We also use our approach to present {\it pure} decoherence and decoherence
accompanied by dissipation in terms of spectral density of reservoir. The
so-called {\it dynamical decoupling} method is discussed in this context.
Finally, we propose {\it minimal decoherence model}, in which the only source
of decoherence is vacuum. We optimize fidelity of quantum information
processing under the trade-off between speed of gate and strength of decoherence.
\end{abstract}

\section{Introduction}
\label{introduction}

In spite of many remarkable results on decoherence in open quantum systems
\cite{cotudac} the problem still remains open involving some conceptual and
interpretational difficulties in the description of dynamical quantum effect.
Recently the problem become crucial in connection with the idea of quantum
computers  \cite{Deutsch,Shor}. The latter  stimulated a huge theoretical and
experimental effort to control evolution of quantum systems. This progress
allows to hope that theoretical obstacles to build quantum computer (QC)
can be overcome. To deal with unavoidable decoherence, the quantum error
correcting codes have been designed \cite{Shor2,Steane} resulting in  the
theory of fault-tolerant (FT) quantum computation
\cite{ShorFT,Dorit,Kitayev,Knill}. In this theory
it is tacitly assumed that decoherence acts {\it independently} of
the structure of the controlled self-evolution of QC (the one
including quantum algorithm and the scheme of error correction). The
self-evolution only {\it propagates} the errors. In fact, decoherence
depends on the kind of dynamics of the system of interest, and
this should be taken into account.

The main purpose of this paper is to provide more complete general analysis
of interrelations between decoherence and controlled evolution,
basing on well established theory of open systems \cite{Alicki-Lendi}.
We develop the dynamical, Hamiltonian  description of decohering quantum
computer, deriving and analysing non-Markovian master equation.
We show that the long-range quantum memory in conjunction with 
self-dynamics of
quantum computer implies highly nonlocal structure of noise.
This has remarkable implications on quantum error correction concept. The 
key idea of the latter is that the noise is local, so that the quantum 
information can be hidden in multi-qubit entanglement. Once we do not 
couple the qubits through quantum gates, the noise is local indeed. However, 
to protect the information against the noise, we need to produce multi-qubit 
entanglement. Due to memory, the reservoir will then ``see'' the 
evolution as highly nonlocal, and after some time it will act 
nonlocally itself. We argue that the long-range quantum memory 
is unavoidable for fundamental reasons: it is exhibited by
interaction with vacuum, which can be never removed. It turns out 
that the memory is relevant even for quantum optics regime, 
despite the fact that the deviation from exponential decay is 
unobservable.

In Refs. \cite{Dorit,Knill} it was noted that FT 
method may not work,  if time or space correlations do not decay 
exponentially. What we show here, is 
that correlations in time cause the noise to follow the evolution of 
the system that is to protect against the noise. 
It follows that FT scheme should be  revisited to take into account 
the memory effect. There is a hope that it is possible, as the 
collective noise attacks due to memory are of special type: they can be 
viewed as a result of propagation backward in time of single-qubit errors.

One should mention, that the long quantum memory is exhibited also by 
phonon environment which characterized stronger coupling than 
electromagnetic  vacuum. This means that  within the program of 
solid state quantum computation the memory effects can become 
practically relevant.


Other programs to avoid decoherence use {\it dynamical decoupling}
\cite{Lloyd}, and {\it decoherence free subspaces} \cite{DFS}. We discuss
the first approach in the context of real physical environment and conclude
that it is hard to find conditions for which the method could be useful.
The decoherent free subspaces approach bases on opposite paradigm
to the one of quantum error correction - it exploits  symmetries
of collective interaction with environment. Even though our approach is
fully general, in this paper we will discuss the independent interaction
case.

We also introduce a {\it minimal decoherence model} of
evolution, within the framework of non-Markovian approach.
The motivation behind the model is to try to take into account
only fundamental obstacles for building quantum computer. That is we
remove any decoherence mechanism that could be, in principle, avoided at
some technology. In the model
the only environment is the vacuum, and there is no internal, natural
self-evolution (the only evolution of the system is constituted by
quantum gates). It follows that {\it there is} some room to optimize the process of
quantum computation within this model.
Subsequently, we obtain time-energy trade-off due
to that fact that to perform a given gate one needs a constant amount of
``action'' that can be split in different ways into time and energy.
Higher energy implies quick gate, but simultaneously, enhances decoherence.
In result it is possible to keep high fidelity at the expense of worse 
scaling. For sequence of $n$ gates, the  physical time needed to 
keep high fidelity scales as $n^{3/2}$.

This paper is organized as follows. In Sect. \ref{sec-preliminaries}
we make some preliminary remarks. Then we develop Hamiltonian
approach in Sect. \ref{sec-Hamiltonian}. We obtain the transition
map (called here error map). Then we pass to the Markovian limit for
time-independent  Hamiltonian and indicate that in non-Markovian stage
the reservoir recognizes structure of levels of the systems. We
discuss possible types of
spectral densities causing decoherence and the problem of high
frequency cut-offs. Subsequently we discuss the dynamical
decoupling method. In Sect. \ref{sec-examples} we present two
examples - QC in
memoryless reservoir, and QC driven by kicked dynamics. We discuss
error path interference effect. In Sect. \ref{subsec-memory} memory
causing by interaction with vacuum is derived. Then we pass to the main
result of the paper (Sect \ref{sec-memory-noise})
deriving and discussing the master equation for decohering QC. We
relate the results to quantum error correction concept.
In Sect. \ref{sec-minimal-model} we postulate {\it minimal
decoherence model}, and provide formula for fidelity within the model.
An example of fidelity for single qubit rotation is presented and
the time-energy trade-off is discussed.

\section{Preliminaries}
\label{sec-preliminaries}

\subsection{Fidelity in interaction picture}
\label{subsec-picture}

We consider the general case of QC interacting with environment $E$.
The initial state of QC is given by $\psi_{t_0}$. The dynamics of QC without
decoherence, is given  by unitary evolution of quantum algorithm
\be
\psi_{t_0}\to\psi_t= U(t,t_0)\psi_{t_0}
\ee
We will call algorithm the total controlled evolution, i.e.
the computation algorithm itself plus the error correction procedures.
Due to interaction with environment, the initial state will evolve into
mixture
\be
\psi_{t_0}\to\varrho_t=\tilde\Lambda(t,t_0) \varrho_{t_0}
\ee
with $\varrho_{t_0}=\psi_{t_0}$. The fidelity of decohered computation
is given by
\be
F_t=\la\psi_t|\varrho_t|\psi_t\ra.
\ee
It is convenient to use interaction picture. Putting
\be
\Lambda(t,t_0)=\hat U(t_0,t)\tilde\Lambda(t,t_0)
\ee
we  obtain
\be
F_t=\la\psi_{t_0}|\Lambda(t,t_0)(|\psi_{t_0}\ra\la\psi_{t_0}|)|\psi_{t_0}\ra.
\ee
Here we used the notation
\be
\hat U(\varrho)=U\varrho U^\dagger,
\ee
where $U$ is unitary operation.

Note that in the interaction picture the quantum algorithm does not compute,
but only ``interacts'' with errors, causing them to spread more and more.
The map $\Lambda(t,t_0)$ describes the net effect of decoherence. As we will
see, in FT model the decoherence is ``decoupled'' from algorithm, so
that the total decoherence amounts to the  faults caused by environment,
which are then propagated by the algorithm.
In Hamiltonian approach, the interaction with environment
will be ``entangled'' with algorithm, due to quantum memory, and no such
simple description will apply.

\subsection{General evolution}
\label{subsec-general}

Through the paper we will deal with equation
\be
{\dt\over \dt t} \sigma_t=\lcal_t \sigma_t +L \sigma_t
\label{eq-ogolna}
\ee
where $\lcal_t$ is some time dependent operator (it will describe
self-evolution), while $L$ is time independent operator
(interaction), $\sigma_t$ is vector.
Let the free evolution of $\sigma$ be given by the operator $\Gamma(t,t_0)$
\be
\Gamma(t,t_0)\sigma_{t_0}=\sigma_t.
\ee
The operator satisfies
\be
{\dt\over \dt t }\Gamma(t,t_0)=\lcal_t \Gamma(t,t_0).
\ee
The total evolution denoted by $\tilde\Lambda(t,t_0)$ satisfies
\be
{\dt\over \dt t }\tilde\Lambda(t,t_0)=\lcal_t \tilde\Lambda(t,t_0) +
L\tilde\Lambda(t,t_0).
\ee
Then the evolution in interaction picture given by
$\Lambda(t,t_0)=\Gamma(t_0,t)\tilde\Lambda(t,t_0)$ we have the following
formal expansion
\ben
&&\Lambda(t,t_0)=\id + \sum_{m=1}^\infty\int^t_{t_0}\dt t_m \ldots
\int_{t_0}^{t_2}\dt t_1 \times \nonumber\\
&&\Gamma(t_0,t_m) L \Gamma(t_m,t_{m-1}) L \ldots L \Gamma(t_1,t_0)
\label{eq-dynamika-inter}
\een
Full evolution is given by
\ben
&&\tilde\Lambda(t,t_0)=\Gamma(t,t_0) + \sum_{m=1}^\infty\int^t_{t_0}\dt t_m \ldots
\int_{t_0}^{t_2}\dt t_1 \times \nonumber\\
&&\Gamma(t,t_m) L \Gamma(t_m,t_{m-1}) L \ldots L \Gamma(t_1,t_0).
\label{eq-dynamika}
\een


\section{Evolution of QC in second order approximation}
\label{sec-Hamiltonian}

In this section we derive reduced dynamics of QC interacting with
environment $R$. Consequently, in eq. (\ref{eq-ogolna})
the operator $\lcal_t$ will be sum of free Hamiltonians  of the QC and $R$,
the operator $L$ --  the interaction Hamiltonian, and $\sigma_t$ -- the wave
function of the total $QC+R$ system.
\subsection{Non-Markovian Born approximation}
\label{subsec-first order}
The Hamiltonian of the total
system is of the form
\be
H=H_{QC}+H_R+\lambda H_{int}.
\ee
Here $H_{QC}$ is time-dependent Hamiltonian of the computer, $H_R$ is
time-independent self-Hamiltonian of the environment, $\lambda$ is
coupling constant, which is
assumed to be small ($\lambda<\kern-.7mm< 1$).
In the following we will remove $\lambda$ from formulas by
incorporating it into interaction Hamiltonian $H_{int}$.
The latter  is	of the form
\be
H_{int}=\sum_\alpha S_\alpha\otimes R_\alpha
\ee
where $S_\alpha,R_\alpha$ are self-adjoint operators. The resulting evolution
of the total system is described by unitary transformation $U(t,s)$ such that
\be
\hat U(t,s)\varrho(s)=\varrho(t).
\ee
We use the notation:
\be
\hat H(\varrho)=[H,\varrho].
\ee
The self-evolution of QC is given by
\be
U_C(t,s)={\cal T} e^{-i\int^t_sH_C(u) \rm d\; u}
\ee
where $\cal T$ is time ordering operator. Finally, the free evolution of the
total system is
\be
\hat U_0(t,s)=\hat U_C(t,s) \otimes e^{-i\hat H_R(t-s)}.
\ee
The initial state of the total system is given by
\be
\varrho(s)=\varrho_C(s)\otimes \omega_R,
\ee
where $\varrho_C$ is the initial state of quantum computer,
while $\omega_R$
is stationary state of environment. Without loss of generality, one can
assume that
\be
\trace R_\alpha \omega_R=0.
\ee
Besides, the state $\omega_R$ commutes with dynamics of environment
\be
[\omega_R,H_R]=0
\ee
We consider lowest order approximation of the evolution, obtaining from
equation (\ref{eq-dynamika-inter})
\ben
&&\hat U(t,s)\simeq \hat U_0(t,s)-i\intlarge^t_s\dt u \hat
U_0(t,u)\hat H_{int}\hat U_0(u,s)-\nonumber\\
&& \intlarge^t_s\dt u\intlarge^u_s \dt w  \hat U_0(t,u)
\hat H_{int} \hat
U_0(u,w)\hat H_{int} \hat U_0(w,s).
\een
Then the evolution of the reduced density matrix
reads as
$\varrho_C(t)=\trace_R(\varrho(t))$ is given by
\ben
&&\varrho_C(t)=\hat U_0(t,s)\varrho_C(s) -
\trace_R
\biggl\{\intlarge^{t}_{s}\dt u  \intlarge^{u}_{s}\dt w
\hat U_0(t,u)\times\nonumber\\
&&\hat H_{int} \hat U_0 (u,w)\hat H_{int}
\hat U_0(w,s)
\varrho_C(s)\otimes \omega_R
\biggr\}.
\een
Subsequently, we find  that $\varrho_C(t)$ can be written in the following
compact form
\ben
&&\varrho_C(t)=\hat U_C(t,s)\biggl(\varrho_C(s)
-{1\over 2}[A(t,s)\varrho_C(s)-\varrho_C(s)A(t,s)]\nonumber\\
&&+\hat\Phi(t,s)\varrho_C(s)-i[h(t,s),\varrho_C(s)]\biggr)
\label{eq-evolution}
\een
The completely positive superoperator $\hat \Phi(t,s)$ is
given by the following formula
\ben
&&\hat \Phi(t,s) \varrho_C(s)=\sum_{\alpha\beta}
\intlarge^{t}_{s}\dt u	\intlarge^{t}_{s}\dt w
\la R_\beta R_\alpha(u-w)\ra_{\omega_R}\times
\nonumber\\
&&S_\beta(u,s)\varrho_C(s)
S_\alpha(w,s)
\label{eq-errormap1}
\een
where $S_\alpha(u,s)=\hat U_C^{-1}(u,s)S_\alpha$,
$R_\alpha(t)=e^{i H_Rt}R_\alpha e^{-iH_Rt}$ are the versions of $S_\alpha$
and $R_\alpha$ evolving in Heisenberg picture according to free evolution;
$\la R_\alpha
R_\beta(t)\ra_{\omega_R}=\trace (\omega_R R_\alpha R_\beta(t))$ is the
autocorrelation function of the environment.
The operator $A(t,s)$ is given by
\be
A(t,s)=A^\dagger(t,s) = \hat \Phi^*(t,s) \id
\ee
where $\hat \Phi^*$ is dual map to $\hat \Phi$.
Explicitly
\be
A(t,s)= \sum_{\alpha\beta}
\intlarge^{t}_{s}\dt u	\intlarge^{t}_{s}\dt w
 \la R_\alpha R_\beta (u-w)\ra_{\omega_R}
S_\alpha(u,s) S_\beta(w,s).
\ee
Finally, $h(t,s)$ describes the Hamiltonian contribution to the dynamics due to
the interaction with environment (Lamb shift, collective Lamb shift, etc.).
In the following we put $h(s,t)\equiv0$ by applying renormalization procedure.
We add to the Hamiltonian $H_{QC}(t)$ the appropriate counter-terms which cancel
the contribution $h(t,s)$ in a given order of perturbation calculus.
Therefore in the following $H_{QC}(t)$ is the full physical Hamiltonian containing all
relevant terms. In this way, when passing to Markovian approximation
for the reservoir at the thermal equilibrium we obtain the Gibbs state corresponding to full  Hamiltonian, as it should be.

One can pass to the frequency domain, putting
\be
\la R_\alpha R_\beta(t)\ra_{\omega_R}=\intlarge^{+\infty}_{-\infty}  R_{\alpha
\beta}(\omega) e^{-i\omega t} \dt \omega,
\ee
\be
Y_\alpha(\omega)=
\intlarge^{t}_{s}  S_\alpha(u,s)
e^{-i\omega u} \dt u.
\label{igrek}
\ee
The function $R_{\alpha\beta}$ is called {\it spectral density}.
We then can write  $\hat\Phi$ as
\be
\hat\Phi(t,s)\varrho_C(s)=\sum_{\alpha,\beta}\intlarge_{-\infty}^{+\infty}
\dt \omega R_{\alpha\beta}(\omega)Y_{\beta}(\omega)\varrho_C(s)
Y^\dagger_\alpha(\omega)
\label{eq-map}
\ee
where we do not write explicitly the dependence on $t$ and $s$  of the
operators $Y_\alpha$.  Denoting $\hat\Phi(t,0)=\hat\Phi_t$, the evolution
in interaction picture can  be
thus written (for $s=0$) as 
\be
\varrho_C(t)=\hat U_0(t,0)
\biggl[\varrho_0 -{1\over 2} \{\hat\Phi_t^*(\id), \varrho_0\} +
\hat\Phi_t\varrho_0\biggr],
\ee
where $\varrho_0=\varrho_C(0)$. The above equation closely resembles
the form of  the generator of the Markovian semi-group
\cite{GKS,Alicki-Lendi}.
In next section we will exhibit Markovian approximation of this formula.

One notes that the crucial element is here the
transition map $\hat\Phi$ (we will call it {\it error map}). Operators 
$S_\alpha$ are errors that can occur during interaction with reservoir 
(cf. \cite{Viola-gen}).
In particular, the map contains propagation of errors which is nothing but
evolution of $S_\alpha$ in interaction picture. The overall error 
is proportional to $\|\hat\Phi\|$, where $\|\cdot\|$ is some norm
(see \cite{Dorit}).

\subsection{Markovian limit for time independent Hamiltonian}
\label{subsec-Markovian}
Let us put $t={\tau\over 2}$, $s=-{\tau\over 2}$, and
\be
H_C=\sum_j\epsilon_j|j\ra\la j|
\ee
where $\{|j\ra\}$ is the orthonormal basis in the Hilbert space of QC.
To find the meaning of the completely positive map $\Phi$, we
will find the behaviour of evolution for long time.
We have
\be
S_\alpha(u,s)=\sum_{\omega_k=\epsilon_j-\epsilon_{j'}}
S_\alpha(\omega_k) e^{i \omega_k u} e^{-{i\over 2}\omega_k\tau}
\ee
with
\be
S_\alpha(\omega_k)=\sum_{j,j':\epsilon_j-\epsilon_{j'}=\omega_k}
|j\ra\la j|S_\alpha |j'\ra\la j'|.
\label{somegak}
\ee
Then the formula (\ref{eq-map}) reads as
\ben
&&\hat\Phi(s,t)\varrho_C(s)=\biggl[\pi\sum_{\alpha,\beta,\omega_k}\intlarge
\dt \omega R_{\alpha\beta}(\omega)\delta^{(2)}_\tau(\omega-\omega_k)
\nonumber \\
&&S_\beta(\omega_k)\varrho_C(s) S_\alpha^\dagger(\omega_k)\biggl]\tau \nonumber\\
&&+\pi^2\sum_{\alpha,\beta}\sum_{\omega_k\not=\omega_l}\intlarge
\dt \omega R_{\alpha\beta}(\omega)
\delta^{(1)}_\tau(\omega-\omega_k)\delta^{(1)}_\tau(\omega-\omega_l)\nonumber\\
&&S_\beta(\omega_k)\varrho_C(s) S_\alpha^\dagger(\omega_k) e^{-{i\over 2}
(\omega_k-\omega_l)\tau}.
\label{eq-map2}
\een
Here we have  used two models of Dirac delta function with
width ${1\over \tau}$
\be
\delta^{(1)}_\tau(x)={\sin \tau x\over \pi x}, \quad
\delta^{(2)}_\tau(x)={\sin^2 \tau x\over \pi x^2 \tau}.
\label{delty}
\ee
Approximately we have $({\delta^{(1)}_\tau})^2\simeq\pi\tau\delta^{(2)}_\tau$.
The second, ``non-resonant'' term of eq. (\ref{eq-map2}) will vanish in the limit of large $\tau$,
as the overlap between two deltas will decrease for large $\tau$.
(This is an alternative form of a well known rotating-wave approximation).
More precisely, one requires $\tau>\kern-.7mm> {1\over \Delta \omega}$, where
\[
\Delta\omega=\min\{|\epsilon_j-\epsilon_{j'}|;\epsilon_j\not=\epsilon_{j'}\}
\]
Finally, we obtain
\be
\hat\Phi(t,s)\varrho_C(s) = \tau\pi \sum_{\alpha,\beta,\omega_k}
R_{\alpha\beta}(\omega_k)
S_\beta(\omega_k)\varrho_C(s) S_\alpha^\dagger(\omega_k)
\label{eq-cpmap-markow}
\ee
Note that the map depends on $t$ and $s$ only through the factor $\tau=t-s$.
Thus ${1\over \tau} \hat \Phi$
is exactly the transition map for
the quantum dynamical semi-group in the weak coupling limit
\cite{Davies,Alicki-Lendi}
\be
\dot{\varrho}_C=
L\varrho_C
\ee
with
\be
L\varrho_C=-i[H_C,\varrho_C]-{1\over 2\tau}\{\hat \Phi^*(\id),\varrho_C\}
+{1\over \tau}\hat\Phi\varrho_C.
\ee
One can see \cite{AlickiPRA} that we can distinguish the initial,
non-Markovian stage, when reservoir is ``learning'' the structure of
the Hamiltonian of the system. This requires time
$\tau>\kern-.7mm> {1\over \Delta \omega}$. Once the structure is recognized,
we have Markovian stage, during which the system is being relaxed towards
the Gibbs state of thermal equilibrium ${1\over Z}e^{-\beta H}$ determined by
$H$.

Suppose now that we have two qubit system, where the components are
coupled to one another:
e.g. the self evolution is to produce some two qubit gate. Then, in
Markovian stage the system relaxes to {\it compound} Gibbs state. Thus,
the noise, after recognizing that the Hamiltonian is compound, becomes
compound itself \cite{generatory}.
We will analyse it in more detail in Sect. \ref{sec-memory-noise}.

\subsection{Decoherence and dissipation in Markovian regime.}
\label{subsec-decoherence-dissipation}

The properties of the decoherence in Markovian limit can be read
from the formula (\ref{eq-cpmap-markow}). We will now analyse the formula
to get first intuition about decoherence.
First of all, one can single out {\it pure decoherence} which is not connected
with energy exchange with environment. The populations of energy levels
are kept, but the phases subject randomization.
The relevant term  is
\be
\pi\tau
\sum_{\alpha,\beta} R_{\alpha,\beta}(0) S_\beta(0)\varrho_C 
S_\alpha^\dagger(0).
\ee
Indeed $S_\alpha(0)$ commutes with $H_C$. Hence this term does not lead to 
transitions between energy levels. Thus the term $R_{\alpha\beta}(0)$  stands 
for the strength of pure dephasing. On the contrary 
the other terms of the sum correspond to the decoherence accompanied by
energy exchange, because the operators $S_\alpha(\omega_k)$ either 
do not exist (if $S_\alpha$ commute with  $H_C$) or describe transitions 
between energy levels.    
Note that if the spectral density $R_{\alpha\beta}$
vanishes for $\omega=0$,
there is no pure decoherence: the decoherence is always accompanied by
dissipation. Assume, for example, that a system does not have
self-Hamiltonian (e.g. spin of free electron). Then there is no dissipation.
If in addition $R_{\alpha,\beta}(0)=0$ then there {\it no} decoherence at all.

\subsection{Important examples of reservoirs}
\label{subsec-reservoirs}

We note that an important characteristic of the interaction with environment
is the shape of the spectral density. A general
property of the heat bath at the temperature $T$ is the following
\be
R_{\alpha\beta}(-\omega)=e^{-{\omega\over kT}}R_{\alpha\beta}(\omega)
\label{KMS}
\ee
In particular, at zero temperature, only nonnegative frequencies
are relevant.
Let us now present three important examples of environments.

\paragraph{Linear coupling to bosonic field}
Operators $R_\alpha$ are given by
\be
R_\alpha=a(\phi_\alpha)+a^\dagger(\phi_\alpha)
\ee

Where $a^\dagger(\phi_\alpha)$, $a(\phi_\alpha)$ are creation and anihillation
operators for fields $\phi_\alpha$, respectively.
The following form of diagonal elements  of $R_{\alpha\beta}$
can be obtained
\be
R_{\alpha\alpha}(\omega)=\intlarge \dt \kvec\delta (\Omega(k) -\omega) (n(k)+1)
|\phi_\alpha(\kvec)|^2
\ee
where $\Omega(k)$ is the energy of the boson ($\Omega(k)={k^2\over 2m}$ or
$\Omega=vk$), $n(k)={1\over e^{\beta \Omega(k)}-1}$.
Typically $R_{\alpha\alpha}(\omega)$ grows like $|\omega|^d$ for small 
$\omega$ and then rapidly falls down  for $|\omega|>\omega_c$
($\omega_c$ is a cut-off frequency). 
%
For electromagnetic interactions in dipol approximation we have
$R_{\alpha,\alpha}(\omega)\approx\omega^3$ for small $\omega$. Similarly for some
cases of phonon interaction. Cut-off parameter $\omega_c$ is
{\it model-dependent} (see, e.g., \cite{Ford}) and characterizes 
the range of validity of model but
does not mean that the frequencies $\omega>\omega_c$ are not influenced
by noise. 
For example, the electromagnetic interaction can be described by the dipol
approximation for ${\omega\over c}=|{\bf k}|\duzomniejsze {1\over r_0}$
where $r_0={e^2\over m_e c^2}$ is classical electron radius \cite{Mesiah}. 
For larger frequencies different model
involving non-bounded (scattering) states of electrons should be taken into
account. The other cut-off for free electrons $\omega_c\simeq 2m_ec^2$ means
only that for larger frequencies the pair production should be taken into
account. In solid state $\omega_c$ is a Debye frequency. For higher
frequencies the coupling to phonons makes no sense and we have to consider
local interactions with the neighbouring atoms or ions and electrons.

\paragraph{Interaction with a dilute gas.}

We use the model of free bosonic or fermionic gas in the low density
approximation
\be
R_\alpha=a^\dagger(\phi_\alpha)a(\phi_\alpha) -
\la a^\dagger(\phi_\alpha)a(\phi_\alpha)\ra
\ee
The spectral density is then of the form
\be
R_{\alpha\alpha}(\omega)= \intlarge\!\!\intlarge \dt \kvec \dt \lvec n(k) |\phi_\alpha(\kvec)|^2
|\phi_\alpha(\bbox{l})|^2\delta(\Omega(k)-\Omega(l)-\omega)
\ee
Typically $R_{\alpha\alpha}(0)>0$ and, again  
$R_{\alpha\alpha}(\omega)$ falls down 
for $|\omega|>\omega_c$ -- a model dependent cut-off frequency. 

\paragraph{Fluctuations of molecular field.}
Influence of a local fluctuating field can be described by a classical
noise  $F_\alpha(t)$. For example for a coloured noise model we have
\be
\la F_\alpha F_\alpha(t)\ra=D e^{-{|t|\over \tau_c}}
\ee
Then the spectral density is Lorentzian 
\be
R_{\alpha\alpha}(\omega)={D\over \omega^2+\tau_c^{-2}}.
\ee
Such a density is relevant for relaxation mechanism in pulsed NMR 
experiments.


\subsection{Discussion of dynamical decoupling method}
\label{decoupling}
From eq. (\ref{eq-cpmap-markow}) we see
that one can gamble with decoherence by changing spectrum of the Hamiltonian
trying to choose frequencies $\omega_k$ for which the spectral density is
small.       In Ref. \cite{Lloyd} the method of {\it dynamical decoupling} was
exhibited. It bases on adding periodic, rapidly alternating term to the
self-Hamiltonian of QC  which averages out the  interaction part, leaving some
room for
controlled evolution of QC. The method is supported by an elegant 
group-theoretic  framework. Here we would like to determine
what types of reservoirs allow to apply such method. By use of a toy
model we will argue, that, unfortunately, typical reservoirs presented in sec.
\ref{subsec-Markovian}
do not allow for dynamical decoupling as proposed in \cite{Lloyd}. 
Consider decoherence in QC with ``bang-bang'' term only, given 
by 
\be
H(t)=H \cos \Omega t 
\ee
where $H$ is time independent, $\Omega$ is large frequency. We need to
derive $Y_\alpha(\omega)$. Using notation of eq. \ref{somegak} we have
\be
S_\alpha(t,0)=\sum_{\omega_k}S_\alpha(\omega_k) e^{i{\omega_k\over \Omega}
\sin\Omega t}
\ee
Expanding the time dependent term into Taylor series and applying Fourier transform we obtain 
\ben
&&Y_\alpha(\omega)=\sum_{\omega_k}S_\alpha(\omega_k)
\bigl[c_0\delta(\omega)+\nonumber \\
&&c_1 {\omega_k\over \Omega}(\delta(\omega-\Omega)+
\delta(\omega+\Omega))+\ldots  \bigr],
\een
where $\delta$ stands for $\delta^{(1)}_\tau$ (see eq. (\ref{delty})), 
$c_{0,1}$ are constants and
we omitted higher harmonics. Putting $Y_\alpha$ into formula 
(\ref{eq-map}) we see that to obtain low decoherence, we need $R_{\alpha\beta}$ (i) to be small for  $\omega\approx 0$ and  (ii) to have cut-off frequency satisfying 
$\omega_c\leq \Omega$. Thus one would need bell-shaped spectral density.
The  bosonic field reservoirs satisfy (i), but, as we have already mentioned,
do not have
physical cut-offs. On the other hand collisonal or 
the coloured noise have cut-offs,
yet allow
for {\it pure decoherence due to nonzero  $R_{\alpha\beta}(0)$}.
The conditions for dynamical decoupling can be in principle met in
QED cavity. Then however, the system loses its fundamental simplicity: the
quantum computer must now include modes of cavity, while the reservoir
become the atoms building the cavity.

There is another aspect of the discussed method, that has not been
investigated so far. Namely, the time dependent Hamiltonian is obtained by
use of external fields that can be described classically (e.g. coherent
light of the laser beam). However, rapidly alternating field
will get entangled with the controlled system (this is called quantum
back-reaction) so it must be treated quantum mechanically. The resulting
disturbance  was evaluated to be weaker than the ``regular'' (say
collisional) decoherence \cite{back-reaction}. However, it seems
that for the bang-bang control, the considered effect can become relevant.

An interesting  version of dynamical decoupling method  was considered in 
Ref. \cite{Agarwal}. Namely, the starting point was  modulate the 
coupling constant in interaction picture rather than self Hamiltonian. 
As a result, in Schr\"oedinger picture, the self-Hamiltonian was of the 
form $\omega H_C \sin{\omega t}$. Thus in addition to modulation,
the level differences of the Hamiltonian were rescaled. Then the 
above analysis does not apply. However, one can show that in the 
cases where the above method works, more elementary strategies 
can protect the system against the reservoir. 
We will discuss it elsewhere.

\section{Examples}
\label{sec-examples}
We will present here two examples. The first one will serve us to introduce
{\it fault map} describing the action of noise with propagation excluded.
In the second one we consider kicked dynamics of algorithm, and illustrate
the difference between quantum and classical memory.

\subsection{Memoryless reservoir: error map and fault map}
\label{subsec-memoryless}
Memoryless reservoir has the  white-noise spectral density. Its auto-correlation
function is $R_{\alpha\beta}(s-t)=R_{\alpha\beta}^0 \delta(s-t)$.
Let us, for simplicity,  assume independent interaction
$R_{\alpha \beta}(\omega)=R_{\alpha} \delta_{\alpha\beta}$.
The resulting error map is given by
\be
\Phi_t(\varrho)=
\sum_{\alpha} R_{\alpha}^0 \int^t_{s}\dt uS_\alpha(u,s)\varrho 
S_{\alpha}(u,s).
\label{eq-errormap2}
\ee
Suppose that $S_\alpha$ are one-qubit operators. According to notation of
Ref. \cite{Dorit} we will call them {\it faults} - these are errors
caused by the reservoir, and should be distinguished from the {\it errors}
resulting from propagation of faults by algorithm (we will denote 
it by $U(t,s)$ in this section).
The propagation of the faults is described by the fact that the error map
involves $S_\alpha$ in Heisenberg picture. Note, that the overall error
is caused solely by one qubit faults and their propagation, averaged
in time. Such  evolution can be described by the following master equation
\be
{\dt \over \dt t} \varrho_t = i[H_t,\varrho_t] +L\varrho_t
\label{eq-master-delta}
\ee
with
\be
L=-{1\over 2}\{\hat\Phi^*(\id),\varrho\} + \hat\Phi(\varrho)
\ee
for $\hat\Phi(\varrho) =\sum_\alpha S_\alpha \varrho S_\alpha$.
Indeed, let equation (\ref{eq-ogolna}) $\sigma_t$ be density matrix of
$QC$, $\lcal_t=-i\hat H_t$ - where $H_t$ is Hamiltonian of
algorithm and  $L$  - the above generator.
Applying the formula (\ref{eq-dynamika})
in first order approximation we obtain
\ben
&&\varrho_t=\hat U(t,t_0)\bigl[\varrho_{t_0} -{1\over 2}
\{A,\varrho_{t_0}\} +\nonumber\\
&&
+\sum_\alpha\int^t_{s}\dt uS_\alpha(u,s)\varrho
S_{\alpha}(u,s)\bigr]=
 \nonumber\\
&& =\hat U(t,t_0)\bigl[\varrho-
{1\over 2}\{\Phi_t^*(\id),\varrho\} +\Phi_t(\varrho)\bigr].
\een
Even though the memoryless case is not especially interesting from our point
of view, the very form of the  master equation
(\ref{eq-master-delta}) is particularly  useful in our context, as the
faults themselves and the propagation are separated from one another. The
propagation is described by the Hamiltonian term, while the faults are described
by the map $\Phi$, which we will call {\it fault map}. We will derive such
master equation in general case in next section. Here, let us only mention
that the ``fault part'' of the master equation is a sum of
one-qubit operators. This means that environment acts {\it locally}, and the
reason for occurring multi-qubit errors is solely due to propagation.

\subsection{Decoherence under kicked dynamics of QC:
Error path interference effect}
\label{sec-kicked}
We will assume that the gates are performed quickly, so that they can be
generated by Hamiltonian with time dependence given by delta function.
This will not work for electrodynamic or phonon
vacuum reservoir,  as due to the behaviour $\omega^3$, any rapid changes in
self-Hamiltonian cause large or infinite contribution. Thus in this case one
needs to work with Gaussian pulses. On the other hand, the kicked dynamics
can be used for Lorentzian spectral density.

Thus we divide computation time $t$ into $N$ pieces of length $\tau$. The
Hamiltonian is given by
\be
H(t)=\sum_{j=1}^N\delta(t-(j-1)\tau)h_j
\ee
where $h_j$ generates $j$-th step of computation. The unitary
evolution is given by
\be
U(t)=\vec\Pi_{j=1}^N\theta (t-j\tau)U(j,j-1)
\ee
with $U(j,j-1)=e^{-ih_j\tau}$ is the $j$-th step of computation
(the group of gates performed at time $j\tau$), $\theta$
is step function. Consequently
\be
U_t=
\left\{
\begin{array}{ll}
U(1,0)& t\in\la 0,\tau)\\
U(2,1)U(1,0)& t\in\la \tau,2\tau)\\
\ldots& \\
U(N,N-1)U(N-1,N-2) \ldots & \\
 \ldots	 U(1,0) & t\in\la(N-1)\tau,N\tau)
\eea
\right.
\ee
Substituting such dynamics into the error map of eq (\ref{eq-errormap1}),
we obtain
\be
\Phi^H=\sum_\alpha\sum_{j,k=1}^N r_{jk}^\alpha S_\alpha^j\varrho S_\alpha^k
\label{eq-discrete-H}
\ee
where $S_\alpha^j=U_j^\dagger S_\alpha U_j$.
For simplicity  we assumed that
$R_{\alpha\beta}=R_{\alpha}\delta_{\alpha\beta}$.
The coefficients $r_{jk}^\alpha$ are given by
\be
r_{jk}^\alpha=\int^{k\tau}_{(k-1)\tau}\!\dt  u
\int^{j\tau}_{(j-1)\tau}\! \dt v R_{\alpha\alpha}(u-v).
\ee
For memoryless case we obtain discrete counterpart of
\be
\Phi^{Ph}=\tau \sum_\alpha  R_\alpha^0
\sum_{k=1}^N S_\alpha^j\varrho S_\alpha^k,
\label{eq-discrete-Ph}
\ee
In both formulas we see propagation of error, this time, in discrete manner:
$S_\alpha^j$ denotes {\it error} resulting in propagation of {\it fault}
$S_\alpha$ within the interval $(0,j\tau)$.
The difference is that  in equation
(\ref{eq-discrete-H}) we have {\it interference} due to quantum memory.
This effect of interference of error can remarkably modify the 
process of error propagation.


The difference is similar to that between separable and
entangled mixed states. We can view the error map as (subnormalized) density
matrix and the Kraus operators as the pure  components of ensemble giving
rise to the density matrix. In general, the noise process can then behave
as classical-like or as quantum stochastic process with time
correlations of the form
\ben
&&\Phi_{cl} \varrho=\sum_{j_1\ldots j_m} p(j_1\ldots j_m) S(t_{j_m}) \ldots S(t_{j_1}) \varrho
S(t_{j_1})^\dagger \ldots S(t_{j_m})^\dagger\nonumber\\
&&\Phi_{qu} \varrho=\sum_{j_1\ldots j_m, i_1,\ldots, i_n} C(j_1\ldots j_m|i_1\ldots i_m)\nonumber\\
&&S(t_{j_m}) \ldots S(t_{j_1}) \varrho
S(t_{i_1})^\dagger \ldots S(t_{i_m})^\dagger,\nonumber\\
\een
respectively.
In the first case we have always mixture of product Kraus operators,
in the second one - we can have a map for which such representation
may be impossible. Indeed, in equation (\ref{eq-discrete-Ph}) the errors occurring in different
times do {\it not} interfere with each other, they are added as in classical probability 
calculus.

However it seems that the crucial point is not the difference between
quantum and classical memory, but its {\it range}. In Sect.
\ref{sec-memory-noise} we will
exhibit remarkable connection between the range of the memory and occurring
multi-qubit faults: the interference caused by long range quantum memory
will give raise to nonlocal noise.

\subsection{Memory of electromagnetic  vacuum reservoir}
\label{subsec-memory}
Here we will calculate and discuss the memory caused by vacuum in free space.
This is important point in our paper, as the vacuum is unavoidable, we cannot
remove it. Of course we can put the QC into cavity, but then, we can
consider the QC plus cavity as the total system, that is again in free space.
We then prefer the latter setup, as it is more fundamental.

Consider the two-level atom  and electromagnetic
field with dipol interaction
\be
H={1\over 2} \omega_0\sigma_3 + \sigma_1 \otimes \bbox{d E}
\ee
The spectral density of reservoir  in temperature $T$ is given by
\be
\bea{lll}
R(\omega)= {8\pi\over 3} \bbox{d}^2 \omega^3
\left(1+ {1\over e^{\hbar\omega/k T}-1}\right)& \mbox{ for }& \omega>0\\
R(\omega)= {8\pi\over 3} \bbox{d}^2 |\omega|^3
\left({1\over e^{\hbar\omega/k T}-1}\right)& \mbox{ for }& \omega\leq0\\
\eea
\ee
We must now calculate autocorrelation function in zero temperature.
Taking inverse Fourier transform we obtain
\be
R(t)={1\over (t+i\epsilon)^4}.
\ee
We see that the memory scales as power of $t$ rather than exponential
function. Then there is no characteristic time. This is due to the fact, that
Coulomb interaction has no characteristic range. Thus the range of vacuum
memory is infinite. Nevertheless, in quantum optics there is a notion of time
of memory, which is the inverse of $\omega_0$ - the transition frequency for
the atom, and one can work with Markovian master equation, within times much
longer than $1/\omega_0$. 

This  does not mean that there exists some characteristic time scale for 
vacuum reservoir. Rather, in the interaction picture, the autocorrelation 
function acquires oscillating term $e^{i\omega_0 t}$ which effectively makes the memory much weaker. 
For optical transitions, the memory has then  almost no observable 
consequences. One could think that the memory problem is not relevant 
in optical implementations of quantum computation. However, it is 
not the case. For two atoms, there are degenerated levels, so  that if 
one wants to 
drive gate coupling these levels, there is no high frequency that would 
diminish the memory effect: the frequency driving the gate and the 
frequency of the transition are the same. Then  the time of 
memory and 
time of controlled evolution are of the same order. The resulting 
non-Markovian effects \cite{Knight75} will be most relevant for many atoms  (as needed 
in quantum computation schemes), as  the number of  degenerate levels 
scales exponentially with number of  atoms, so that it will be 
hard to perform computation without use of the gates suffering 
from memory effects.


\section{Quantum long range memory implies nonlocal structure of noise}
\label{sec-memory-noise}

\subsection{Master equation}
\label{subsec-master}

As we have seen, to discuss the structure of noise  without propagation
effect, we need master equation of the form
\be
\dot{\varrho}= -i  [H_{QC}(t), \varrho ] + \hat L_t \varrho.
\ee
we will now derive such equation in second order approximation. To this end,
we differentiate the equation (\ref{eq-evolution}) obtaining
\ben
&&\dot\varrho(t)= \dot{\hat U_C}(t,s) \varrho(s)+
\dot{\hat U_C}(t,s)\left[ -{1\over 2}
\{\hat\Phi^* \id, \varrho(s)\} + \hat \Phi \varrho(s)\right ] +\nonumber \\
&& \hat U_C(t,s) \left [
-{1\over 2}\{\dot{\hat\Phi}\id, \varrho(s)\} + \dot{\hat \Phi} \varrho(s)
\right].
\een
We make first step of approximation by removing the second term.
Differentiating $\hat \Phi(t,s)$ and putting $s=0$ we obtain
 \be
 \dot\varrho =-i[H_{QC}(t), \varrho] - {1\over 2} \{\hat\Upsilon^*_t \id, \varrho\}
 +\hat\Upsilon_t\varrho
 \ee
where $\hat\Upsilon_t=\hat\Upsilon(t,0)$ is given by
\be
\hat\Upsilon(t,s) (\cdot)= \sum_{\alpha\beta} (\Upsilon_{\alpha\beta}
(\cdot) S_{\alpha} + S_\beta (\cdot) \Upsilon^\dagger_{\alpha \beta})
\ee
with
\be
\Upsilon_{\alpha\beta}=\int^t_s \dt u  R_{\alpha\beta}(u-t) S_{\beta}(u,t)
\label{eq-fault}
\ee
On the right hand side of  the equation we have density matrices
evolving according to self-Hamiltonian of the system.
The second step of approximation is then to replace them with the
ones subjected to full evolution. (Better justification of such
integral differential master equation can be obtained within cumulant
expansion method, see e.g. \cite{Kubo}).

Now, the decoherence is essentially characterized by the map $\hat\Upsilon_t$
which we will call ``fault map'' in analogy to the error map $\hat\Phi_t$.
The latter one characterizes error, i.e. the net effect of the
``attacks'' of noise (faults) and their propagation. The present map
$\hat \Upsilon_t$ describes solely the  action of noise. The propagation
will be due to the Hamiltonian term in the above master equation.

Let us now discuss how the form of $\Upsilon_t$ depends on memory. The crucial
term is, of course, $\Upsilon_{\alpha\beta}$. In the case of no memory
(and, for simplicity, independent decoherence), we obtain
$\Upsilon_{\alpha\beta}=\delta_{\alpha\beta} S_{\alpha}$ which is compatible
with Sect. \ref{subsec-memoryless}.Assuming that each $\alpha$ denotes
different qubit, we obtain that decoherence is local, and multi-qubit errors
are solely due to propagation. We can then write the generator in the form
\be
L_t=\sum_\alpha L_t^\alpha.
\ee
where $L_t^\alpha$ operates only on $\alpha$th qubit. If we admit exponentially
decaying memory  (e.g. coloured noise) with characteristic time $\tau_R$
comparable with the time of performing single computational step,
but short in comparison with time between gates, there will be some time for
reservoir to learn something about the structure of two qubit Hamiltonians,
so that the noise will be a sum of two qubit superoperators
\be
L_t=\sum_{\alpha\beta} L_t^{\alpha\beta}.
\ee
In both cases decoherence is local. The possible faults are, roughly speaking,
symbolized by $L_\alpha^t,L_{\alpha\beta}^t$ and they couple at most
two qubits (in general, they involve so many qubits, as the used gates do).

Suppose now, that the reservoir has long range memory.
Still, in each step of computation, there are two-qubit gates.
The environment learns about their
structure a bit. The effect {\it accumulates} due to memory, so that after long
time, the noise is strongly nonlocal: the generator $L_t$ involves multiqubit
operators as it keeps the record of all the history of self-evolution (i.e.
algorithm). The decoherence becomes as ``entangled'' as the quantum algorithm
is entangling. Note, that the effect does not depend on
the time of performing gates. For short gates, the reservoir has short time
to distinguish the temporary levels (driving the gate), but  they are
easily distinguishable, as the energy differences are high. If instead 
the energy  differences are low, then
the time must be longer. This is the result of time-energy uncertainty
principle. Thus, the relevant parameter is here ``action'' i.e. product of
time of gate, and strength of applied Hamiltonian (energy). To perform the
gate, the needed amount of action must be always of order of 1.
Thus the general structure of noise does not depend very much
on the way the gates are performed. In Sect. \ref{sec-minimal-model}
we will see that the strength of the decoherence does depend on it, so 
that we have the mechanism of possible optimization.

\subsection{Quantum memory and error correction}
Let us now discuss the implication of the above results on quantum error
correction (QEC) method \cite{Steane,Shor2,ShorFT,Dorit}. The basic idea
of QEC rests on two assumptions:
(i) the environment acts locally (ii) the environment acts independently
of self-evolution of the system. If the above assumptions are satisfied, it is
indeed possible to protect the quantum information against decoherence
by encoding it into highly entangled multiqubit states. In literature
the assumption (i) was made explicitly, while the assumption (ii)
was tacit.
Thanks to assumption
(i) such nonlocally encoded information will not be affected by decoherence
(acting locally). The assumption (ii) ensures that the self-evolution will
not modify the local structure of decoherence. As we have argued, the latter
assumption is never satisfied. If there is short range memory, the
dependence on self-evolution is irrelevant, and QEC still applies. However,
the interaction with vacuum, which is unavoidable, introduces long-range
quantum memory, which causes environment to be rather malevolent: it traces
self-evolution, and the more entangled the latter is, the more nonlocal
the faults become caused by noise. On the other hand, the self
evolution {\it must}
be entangling, since it is to hide the quantum information just into
entangled states. Thus, we have something a bit analogous to the Lenz rule: the
structure of environment attack is proportional to the structure of
the evolution that is to protect against it. It seems to be  a rather
pessimistic result,
as it implies, that the very idea of {\it active} protection against
environment is fundamentally inadequate, as far as {\it long} quantum
computation is concerned.

However, the conclusion need not be so pessimistic. First of all,
decoherence caused by vacuum is small in comparison
with other sources of decoherence, e.g. dephasing due to collisions
(pressure decoherence), coloured noise or thermal noise. It seems that
QEC (and hence FT method) can be used to fight with the latter sources of
decoherence, within the time window in which the vacuum-memory
effect can be neglected. Strictly speaking other types of reservoirs also
have long tails in memory, however, the dominant part decays exponentially.
Thus FT could be certainly performed within the time regime satisfying 
the following
conditions: (i) {\it the memory is exponentially decaying}, (ii) {\it
the memory majorize vacuum memory}. This regime may prove sufficient
to perform some quantum computational tasks.

There is also another point to address \cite{Steane}.  The 
claim that lack of memory is sufficient for FT does not mean that 
it is in general impossible to improve FT  against some kind of 
memory. As a matter of fact, the mechanism producing multi-qubit 
faults we described has  some features that may be used to defend 
FT scheme. Namely, as seen from formula (\ref{eq-fault}), the multi-qubit 
fault at some place $\alpha$ arise from {\it backward} propagation 
(with decreasing strength) of single qubit fault $S_\alpha$.  
Now since FT method is able to fight with propagation forward, it cannot 
be excluded that it will be robust against faults correlated in such a way.
However it goes beyond the scope of this paper.

\section{Minimal decoherence model for quantum controlled systems}
\label{sec-minimal-model}

In this section we will postulate {\it minimal decoherence model}.
We then derive formula for fidelity of quantum computation within
the model.
We shall consider a class of controlled open systems for which $H_C(t)\not=0$
only during computation of subsequent gates.
The spectral density $R_{\alpha\beta}$ satisfies
\be
\intlarge^\epsilon_{-\epsilon}{\spec(\omega)\over \omega^2} \dt \omega
<\infty \quad \mbox{ for a some } \epsilon>0.
\ee
This excludes reservoirs with
any pure decoherence effects, such as gas or coloured noise. Our
motivation is that we would like
to investigate only fundamental  obstacles to build quantum computer.
The pure decoherence is not such a one, as
in principle one could isolate the quantum computer from any influence
of the above kind (even though it might require technology that will
never be achieved). Since obtaining arbitrarily low temperature is also
a matter of technology, we can work with the only environment being vacuum.
 We removed time-independent part of
self Hamiltonian, as it will only cause additional dissipation.

Example of the system could be n well separated spins-$1\over 2$ which
interact with an electromagnetic field and possibly with phonons. The
controlled Hamiltonian $H_C(t)$ can be realized by switching external magnetic
fields and suitable interaction between spins.
Another example is atomic degenerate metastable level, with Stark of Zeeman
splitting used for controlling.  Of course it may be impossible
to keep possibility of controlling interactions (a cornerstone of a quantum
computer) and remove the influence of pure decoherence. However, as said, we
assume the most optimistic case, and keep only the obstacle that cannot
be removed for fundamental reasons. By removing time-independent self
Hamiltonian, we obtain that there is no decoherence on qubits that
are not active. Thus decoherence will be caused by the vacuum
solely due to gate operations that require to split energy levels during
the time of performing operation. In the case
of independent interactions of qubits with vacuum, this temporal difference
of energy levels is made of use by the vacuum which  forces the
upper level to relax by spontaneous emission. In the case of collective
interaction (Dicke limit) this is not the case, where one has decoherence-free subspaces 
\cite{DFS}.

\subsection{Fidelity in minimal decoherence model}
\label{subsec-fidelity-minimal}

We put $\varrho_C(s)=|\psi\ra\la\psi|$ and $s=-\infty$, $t=+\infty$. The
fidelity of computation is given by
\be
F=1-\delta=\la U_C(t,s)\psi|\varrho_C(t)| U_C(t,s)\psi\ra
\ee
We will denote $S_\alpha(t)=S_\alpha(-\infty, t)=\hat
U^{-1}_C(t,-\infty)S_\alpha$.
To avoid infinite contribution from the interaction, we will assume
that the interaction is adiabatically switched off at infinity
(the standard method in quantum field theory \cite{Bog}). Thus
 put into formula (\ref{eq-map}) the following
operators
\be
Y_\alpha(\omega)=
\intlarge^{\infty}_{-\infty} \hat S_\alpha(t) e^{-\epsilon |t|}
e^{-i\omega t} \dt t.
\label{eq-X}
\ee
Performing partial integration  and then passing to  the limit
$\epsilon\rightarrow 0$ we obtain
\be
Y_\alpha(\omega)={1\over i\omega}\intlarge^\infty_{-\infty}\dt t e^{-i\omega t}
\biggl({\dt\over \dt t} S_\alpha(t)\biggr)={1\over \omega} X_\alpha(\omega).
\ee

We then obtain the following formula for error $\delta$
\ben
&&\delta= \sum_{\alpha,\beta}\int^{\infty}_{-\infty}{\dt \omega\over
\omega^2} \spec(\omega)\biggl\{\la\psi|X_\alpha^\dagger(\omega)
X_\beta(\omega)|\psi\ra-\nonumber\\
&&\la\psi|X_\alpha^\dagger(\omega)|\psi\ra\la\psi|
X_\beta(\omega)|\psi\ra\biggr\}
\label{eq-error}
\een
\subsection{Quantum information processing in minimal decoherence regime}
\label{subsec-single-qubit}

We consider the simplest quantum gate: rotation of single qubit by an angle
$\alpha\in[0,2\pi]$, in the case of two-level atom coupled to electromagnetic
field in zero temperature. The time dependent Hamiltonian will be of the
form $H_C(t)={1\over 2} f(t)\sigma_z$, with
\be
\int_{-\infty}^\infty f(t)\dt t = \alpha,\quad
\lim_{t\rightarrow \pm \infty}f(t)=0.
\ee
The interaction Hamiltonian is $H_{int}=\sigma_x\otimes \phi$, and
the spectral density, as discussed earlier, is
\be
R(\omega)=
\left\{\begin{array}{l}
R_0\omega^3\quad \mbox{ for } \omega\geq 0\\
0\quad \mbox{ for } \omega< 0
\end{array}\right.
\ee
The full unitary transformation is given by
\be
U_C=U_C(\infty,-\infty)=e^{-{1\over 2}\alpha\sigma_z}
\ee
The parameter $\alpha$ is the amount of ``action'' needed to perform
the transformation. Roughly speaking, $\alpha$ is product of energy
pumped into system during performing gate and the time of operation.
In the case of non-rectangular pulse shape, the time is given by the
width of pulse.
Typically, if the energy is low, then accompanied dissipation is small, however
time of the operation is long (while in quantum computation, one would prefer
short time gates). Adding more energy, we obtain short time, but
simultaneously, enhance decoherence, causing loss of fidelity.
Let us now calculate  the fidelity. We have
\be
X(\omega)=F_-(\omega)\sigma^-+F_+(\omega)\sigma^+,
\ee
where 
\be
F_{\pm}=\int^\infty_{-\infty}\dt t e^{-i\omega t} f(t) e^{\pm i\phi(t)}
\ee
with $\phi(t)=\int^t_{-\infty} f(u)\dt u$. From the formula (\ref{eq-error})
one easily finds that  any rapid variation of $f(t)$ will result in  large
error. Indeed, the changes produce long tails of the
Fourier transforms  $F_{\pm}$, which integrated with $R_0\omega$  for
$\omega\geq 0$ give poor fidelity. In order to minimize this effect, we choose
Gaussian shape of the pulse
\be
f(t)={\alpha\over \sqrt{2\pi} t_C} e^{-{1\over 2}(t/t_1)^2}
\ee
where $t_1$ can be taken as the time of performing a single gate. Using approximate
formula $\phi(t)\simeq f(0)t+{\alpha\over 2}$ one finds
\be
|F_{\pm}(\omega)|=\alpha\exp\bigl\{ -{1\over 2} (\omega t_C\pm \alpha{1\over
\sqrt{2\pi}})^2\}.
\ee
Consequently the error can be estimated as follows
\be
\delta \simeq (\la\psi|\sigma^+\sigma^-|\psi\ra-|\la\psi|\sigma^+|\psi\ra|^2)
\alpha^2 R_0 t_1^{-2}
\ee
Taking the average over possible input states $\psi$ we get
\be
\delta_1 \sim  R_0\alpha^2 t_1^{-2}
\ee
up to a constant factor.

We execute ``algorithm'' with action $A$ in $n$ steps (gates)
i.e. $\alpha=A/n=O(1)$. 
 It turns out that the formula
(\ref{eq-error}) is additive with respect to composition of gates.
To see it consider $n$ gaussian pulses  of width $t_1$ separated with 
time $\tau=mt_1$, where $m\duzowieksze 1$. This is smooth realization of kicked dynamics considered earlier. 
One finds 
\be
X_\alpha(\omega)=\sum_{j=1}^n e^{-i\omega_j \tau} \hat U^\dagger(j\tau,0)
Y_{\alpha}^j(\omega) 
\ee
where 
\be
Y_{\alpha}^j(\omega)=\int^{\tau/2}_{-\tau/2} e^{-i\omega t}{\dt \over \dt t}
\hat U^\dagger(t,j\tau) S_{\alpha}.
\ee
Now, we have, for example
\ben
&&\la\psi| X^\dagger_\alpha(\omega) X_\beta (\omega)|\psi\ra=
\nonumber\\
&&\sum_{j,j'} e^{-i\omega \tau(j'-j)}
\la U(j\tau,0)\psi| {Y_\alpha^j}^\dagger(\omega) Y_\beta^j(\omega)|
 U(j\tau,0)\psi\ra. 
\een
Taking large separation between gates, we obtain that the terms involving $j\not=j'$ are averaged out by rapid phase rotation. The remaining 
term is nothing but the sum of single gate terms.

Consequently, for $n$ pulses with average error $\delta_1$ the total error
is given by
\be
\delta _n\sim n\delta_1
\ee
Suppose now that we would like to keep the total error below some threshold
$\epsilon$  then we obtain the following estimation for the time of
computation $t_C=n(m+1)t_1$ (where $t_1$ is the time of single pulse)
\be
t_C\geq {1\over \sqrt \epsilon} R_0^{1\over 2}(m+1) n^{3/2}.
\ee

In conclusion, it is possible to keep high fidelity, at the expense 
of worse scaling of the physical time $t_C\sim n^{3/2}$ in comparison 
with ``algorithmic'' time  $t_{alg}\sim n$.

Finally, one should mention, that our result is not restricted to 
the simple one-qubit system.  If one takes two qubits, and some two qubit 
gate, the reasoning will be  similar. If the system is K-qubit, 
(still with, say, at most two-qubit gates), the only difference 
will be the need of scaling fidelity per one qubit as $1/K$
to get high fidelity of the total final K-qubit state. This can be achieved 
by further slowing down the gates. Finally, to keep high fidelity of quantum computer in vacuum 
we need physical time to scale as $t_C\sim n^{3/2}\sqrt{K}$ or 
$t_C\sim n \sqrt v$ where 
$v=nK$ is complexity  of the  problem (or volume of the algorithm) i.e. 
number of steps times number of qubits needed to run the algorithm.


\section{Concluding remarks}
\label{discussion}

We have developed dynamical description of decohering quantum computer. We
have  obtained dependence of the structure of decoherence
on quantum memory of the reservoir. The nonlocal structure of decoherence
implied by vacuum memory suggests, that long quantum computation at
constant error rate is impossible.
Also the  high frequency ``bang-bang'' control is not useful for
typical reservoirs. Instead,
one should deal with short time quantum computation by use of low  frequencies.
  This of course requires optimization of the kind we performed in the
last section.   Finally, we have designed the most optimistic {\it minimal
decoherence model}, and provided fidelity formula, in first order approximation.
Using it we have shown how the time-energy trade-off influences
scaling of physical time of computation.


In general we have argued that in description of quantum computing the 
Markovian approach fails, as far as linear coupling to boson field 
is considered. The effects are relatively small for coupling to 
electromagnetic vacuum (though certainly much stronger than the 
non-observable non-exponential decay of radiating atom). 
Our results can have practical
meaning for the systems interacting with phonons, like e.g. quantum dots
(cf. \cite{DiVincenzo}), where the coupling is much stronger. 
Then the memory effects can be of similar order as other 
sources of decoherence. There are also less fundamental but practical 
sources of memory such as e.g. heating mechanism in ion traps \cite{Knight-priv}.

Finally, we believe that  our non-Markovian dynamical description 
of quantum computing will have some  meaning for 
future practically useful implementations of quantum computer involving 
large  number of qubits and relatively long time of computation.




The authors are grateful to Martin Plenio for valuable discussion, 
John Preskill, Andrew Steane,  Lorenza Viola for feedback to the first 
version of the paper, and Peter Knight and Maciej Lewenstein for 
helpful comments. 
R.A. is supported by Polish Committee for
Scientific Research, contract No. 2 P03B 042 16.
M. H., P. H. and R. H. are  supported by Polish Committee for
Scientific Research, contract No. 2 P03B 103 16, and the
UE project EQUIP, contract no. IST-1999-11053. 


\begin{references}
\bibitem{Steane}
We are indebted to Andrew Steane and John Preskill for discussion 
on this matter.
\end{references}
\end{document}